\documentclass[11pt]{article}

\usepackage{color}

\usepackage{authblk}
\usepackage{url}
\usepackage{graphicx}
\usepackage{graphics,epsfig}
 \usepackage{amssymb}
 \usepackage{ulem}
 \usepackage{multirow}
  \usepackage{color}
  \usepackage{subfigure}
  \usepackage{subfig}
 \usepackage{amsmath}
\usepackage[margin=1in]{geometry}
\usepackage[numbers,sort&compress]{natbib}
\usepackage{enumitem}
\usepackage{comment}

\newcommand{\bfl}{\begin{flushleft}}
\newcommand{\efl}{\end{flushleft}}
\newcommand{\bea}{\begin{eqnarray}}
\newcommand{\eea}{\end{eqnarray}}
\newcommand{\be}{\begin{equation}}
\newcommand{\ee}{\end{equation}}
\newcommand{\ben}{\begin{enumerate}[itemsep=0pt,parsep=0pt]}
\newcommand{\een}{\end{enumerate}}
\newcommand{\bi}{\begin{itemize}}
\newcommand{\ei}{\end{itemize}}

\newcommand{\cms}{~\mbox{cm}^3/\mbox{s}}
\newcommand{\gev}{~\mbox{GeV}}

\newcommand{\sgm}{\sigma}

\renewcommand{\em}{\it}

\def\nn{\nonumber}
\def\bec{\begin{center}}
\def\eec{\end{center}}
\def\beq{\begin{equation}}
\def\eeq{\end{equation}}
\def\fr{\frac}

\title{Supersymmetry, Nonthermal Dark Matter  and Precision Cosmology }
\author[1]{Richard Easther}
\author[2]{Richard Galvez}
\author[2]{Ogan \"Ozsoy}
\author[2,3]{Scott Watson}
\affil[1]{Dept of Physics, University of Auckland,  Private Bag 92019, Auckland, New Zealand}
\affil[2]{Dept of Physics, Syracuse University, Syracuse, NY 13244, USA}
\affil[3]{KITP, University of California
Santa Barbara, CA 93106-4030 USA}

\date{\today}

\begin{document}

\maketitle

\begin{abstract}

Within the  Minimal Supersymmetric Standard Model (MSSM), LHC bounds suggest that scalar superpartner masses are far above the electroweak scale.  Given a high superpartner mass, nonthermal  dark matter  is a viable alternative to WIMP dark matter  generated via  freezeout. In the presence of moduli fields nonthermal dark matter production is associated with a long matter dominated phase, modifying the spectral index and primordial tensor amplitude relative to those in a thermalized primordial universe.   Nonthermal dark matter can have a higher self-interaction cross-section than its thermal counterpart, enhancing astrophysical bounds on its annihilation signals.   We  constrain the  contributions to the neutralino mass from the bino, wino and higgsino using existing astrophysical bounds and  direct detection experiments for models with nonthermal neutralino dark matter. Using these constraints we quantify the expected change to inflationary observables resulting from the nonthermal phase.  
\end{abstract}
\thispagestyle{empty}

\newpage
\tableofcontents

\section{Introduction}
Cosmological observations allow us to determine the geometry, composition and  age of the universe  with great accuracy, and to tightly constrain the primordial perturbation spectrum.   Big Bang Nucleosynthesis (BBN)  and  the  recently revealed cosmological neutrino background imply that  the universe was thermalized at MeV scales.  Further, the correlation between temperature and E-mode polarization anisotropies  in the Cosmic Microwave Background (CMB) gives strong evidence that  primordial perturbations  were laid down before recombination.

Standard Model  physics cannot generate the primordial perturbations, drive baryogenesis or supply the dark matter content of the universe. Consequently,  key  processes occur at very high energies during the {\em primordial dark age\/} in which the universe is dominated by physics beyond the Standard Model. This period   is   weakly constrained, given our ignorance of the underlying physics.  Crucially, while the neutrino background and BBN require that the universe was thermalized at MeV scales, it need not be thermalized at higher energies.  The equation of state during the primordial dark age determines the expansion rate, and thus the rate at which modes (re)enter the horizon, modifying the observed power spectrum if the spectral index, $n_s$,  is not strictly scale-invariant \cite{Liddle:2003as,Kinney:2005in,Peiris:2008be,Adshead:2010mc,Mortonson:2010er,Easther:2011yq,Norena:2012rs,Martin:2010kz,Martin:2006rs}.  This issue has primarily been discussed in the context of inflation, but  it arises in any mechanism  generating perturbations well beyond Standard Model scales.  

If the primordial universe is thermalized, massive long-lived particles may {\em freeze-out\/} with a final abundance determined primarily by their mass and annihilation cross-section \citep{Bertone:2004pz}. This is the basis of thermal WIMP\footnote{Weakly Interacting Massive Particles} dark matter, which assumes a weak-scale cross-section, $\sigma$ and $\langle \sigma v \rangle_{th} \simeq10^{-26}$ cm$^3/$s, where  $v$ is the typical velocity.  Alternatively,   {\em nonthermal\/} dark matter is produced via the decay of heavier particles  into a long-lived final state and does not  require thermal equilibrium~\cite{Chung:1998rq,Fornengo:2002db,Pallis:2004yy,Gelmini:2006pw,Gelmini:2006pq} (for a review see \cite{Watson:2009hw}, and for recent related work \cite{Allahverdi:2013tca,Hooper:2011aj,Kelso:2013paa}).   Dark matter models are constrained by both direct detection experiments and searches for  astrophysical signals generated by their annihilation products. Nonthermal dark matter can have a higher self interaction cross-section than thermal dark matter so astrophysical signals are potentially stronger for these scenarios, particularly  in indirect experiments such as FERMI and AMS-2 \cite{Gelmini:2008sh,Grajek:2008pg,Grajek:2008jb,Dutta:2009uf,Kane:2009if,Sandick:2011rp}.

Simple supersymmetric (SUSY) versions of the Standard Model face considerable pressure from LHC data, but SUSY remains a candidate symmetry of high energy physics and SUSY models provide a wide range of dark matter candidates.  In the Minimal SUSY Standard Model  (MSSM),  LHC data requires large scalar superpartner  masses of $10$ TeV or more \cite{Feng:2013tvd} while the dark matter species can be lighter. This a situation that naturally leads to nonthermal dark matter production  (e.g. \cite{Douglas:2012bu,ArkaniHamed:2012gw}) and  a similar argument applies to  anomaly mediated SUSY breaking \cite{Moroi:1999zb}.  Further, a primordial nonthermal phase may be generic in SUSY models with a high energy completion in the presence of gravity \cite{Acharya:2009zt,Watson:2009hw}.  This was seen explicitly  in the {\em G2-MSSM} \cite{Acharya:2008bk}, and later generalized to models  with {\em strong} moduli stabilization \cite{Dudas:2012wi, Evans:2013lpa}. In many cases nonthermal dark matter production is naturally favored in these scenarios.    

Given LHC bounds on the SUSY spectrum in the MSSM, cosmological constraints -- while indirect -- are key to explorations of the SUSY parameter space at higher energies.  In this paper we explore nonthermal dark matter production in the MSSM,  quantifying the extent to which the  nonthermal phase changes expectations for inflationary observables. For nonthermal production, the cross-sections are often larger than typical for thermal dark matter, increasing the sensitivity of astrophysical searches for  dark matter decay products.  In particular,  we discuss constraints on the mass of  neutralino dark matter and the allowed contributions to the neutralino mass from the bino, wino and higgsino.  

The paper is organized as follows. In Section 2 we review uncertainties in inflationary observables derived from the unknown post-inflationary equation of state. In Section 3, we summarize nonthermal dark matter phenomenology and the associated  expansion history.  In Section 4, we explore the post-inflationary expansion history of the universe in an MSSM model with SUSY breaking above the TeV scale, and show how this is constrained by existing and future constraints from dark matter experiments.  In the final section we conclude.

\section{CMB Uncertainties from the Post-Inflationary Expansion}\label{N&Obs}
 
 To determine the predictions of a specific inflationary model\footnote{We focus on inflation, but our arguments apply to any mechanism which generates perturbations on super-Hubble scales with a power spectrum whose spectral index in not strictly scale-invariant.} we match the comoving wavenumber $k$ to the instant it exits the Hubble horizon~\cite{Liddle:2003as,Kinney:2005in,Peiris:2008be,Adshead:2010mc,Mortonson:2010er,Easther:2011yq,Norena:2012rs,Martin:2010kz,Martin:2006rs}.  This occurs when  $k=a_{k}H_{k}$ where $H$ and $a$ denote the Hubble parameter and scale factor respectively, and a subscript $k$ labels values at horizon crossing.  We define $N$, the number of e-efolds before the end of inflation, 
\beq\label{nk_equation_0}
  e^{N(k)} \equiv \frac{a_{end}}{a_{k}},
\eeq
where $a_{end}$  denotes the scale factor at the end of inflation and rewrite $N(k)$ as
\beq\label{nk_equation_1}
  N(k) =\ln \left( \frac{H_{k}}{H_{end}} \right)-\ln \left( \frac{k}{a_0 H_0} \right)  + \ln \left( \frac{a_{end} H_{end} }{a_0 H_0} \right),
\eeq
where $(a_0H_0)^{-1}$ is the value of the co-moving Hubble radius today \cite{Adshead:2010mc}.
The first term in equation \eqref{nk_equation_1} can be determined for any specific model, while the final term depends on the post-inflationary expansion history of the universe.  We assume single-field slow-roll inflation for the purposes of illustration and characterize the post-inflationary expansion by an effective equation of state $w$.  One finds the matching equation~\cite{Liddle:2003as}
\bea\label{matching}
  N(k,w) &\simeq& 71.21
        - \ln \left( \frac{k}{a_0H_0} \right )
        +\frac{1}{4} \ln \left( \frac{V_k}{m_p^4} \right) +\frac{1}{4} \ln\left( \frac{V_k}{\rho_{end}} \right) 
        + \frac{1-3w}{12\left( 1+w\right)} \ln\left( \frac{\rho_{{r}}}{\rho_{{end}}} \right),
\eea
where $\rho_{end}$ is the value of the energy density at the end of inflation, $V_k$ is the inflaton potential as the $k$th mode leaves the horizon, and $\rho_r$ is the energy density at which the universe is assumed to become  thermalized.  The first two terms in (\ref{matching}) are model independent. For GUT scale  inflation the third term is roughly $-10$. The   fourth term is typically order unity given that the value of the inflaton potential necessarily evolves slowly as inflation proceeds.   Finally, if the universe thermalizes promptly  the last term is negligible, and we recover the familiar result that  $50 \lesssim N \lesssim 60$ for modes contributing to the CMB.

If $\rho_{end}^{1/4} \gg \rho_{r}^{1/4}$  and $w\ne 1/3$ then $N$ differs from its benchmark value\footnote{In models that make an explicit prediction for $\rho_{r}$ we can  insert this value into equation~(\ref{matching}). If the post-inflationary thermal history is  unknown  $\rho_r^{1/4}$ is  the energy scale by which thermalization is required to have occurred \cite{Easther:2011yq}.} by 
\be \label{deltaN}
 \Delta N = \fr{1-3w}{12(1+w)}\ln \left( \fr{ \rho_{r} }{ \rho_{end}} \right ), 
\ee
where  $\rho_{\rm{end}} = 3 V_{\rm{end}}/2$ for a given potential. If $w<1/3$,  $\Delta N$ is negative, since  $\rho_{r} < \rho_{end}$.   

The equation of state during the primordial dark age  induces uncertainties in  inflationary predictions for the scalar tilt and tensor-to-scalar ratio $n_s$ and $r$. The uncertainty in $n_s$ is clearly associated with the running $\alpha_s= dn_s/d\ln k$ and to lowest order in slow roll  \cite{Hoffman:2000ue,Schwarz:2001vv,Kinney:2002qn}
\bea\label{deltanr}
\nn \Delta n_s & = & \left. (n_s-1)\left[-\fr{5}{16}r-\fr{3}{64}\fr{r^2}{n_s-1}\right]\right| \Delta N,\\
\Delta r& = & \left. r \left[(n_s-1)+\fr{r}{8}\right] \right| \Delta N.
\eea 
The resulting fractional  uncertainties  $\Delta r/r, ~\Delta n_s/|n_s-1|$ in these  observables can be substantial \cite{Kinney:2005in,Adshead:2010mc}. In particular, the theoretical uncertainty in $n_s$ can be comparable to the precision with which it is measured by Planck \citep{Ade:2013rta}.   Our primary focus is the implications for  $w$ and $\rho_{r}$ of MSSM scenarios with nonthermal dark matter, which will  lead to tighter predictions for the primordial spectrum of specific inflation models.  

\section{Thermal and Nonthermal Dark Matter \label{sec2}}

In the early universe, the  density in WIMPs relative to the critical density at freeze-out is~\cite{Kolb:1990vq}
\be \label{totalthermaldm}
\Omega_{dm}h^2\simeq 8.63 \times 10^{-11} \left(
\frac{m_X }{g_*^{1/2} \langle \sigma v \rangle T} \right) \,  \mbox{GeV}^{-2}.
\ee 
where $m_X$ is the dark matter particle's mass, $\langle \sigma v \rangle$ is the {\em total} thermally averaged cross-section, $g_*$ and $T$ are the number of relativistic degrees of freedom and temperature at freeze-out and $h$ is the present Hubble parameter in units of $100$~km/s/Mpc.  If the universe  is thermalized,  freeze-out occurs at $T_f \simeq m_X/20$ and $g_* \sim 100$, assuming the effective number of degrees of freedom is similar to that of the Standard Model \cite{Beringer:1900zz}.
The abundance simplifies to
\be \label{thermal_abundance}
\Omega_{dm}^{therm}h^2\simeq 0.12  \left( \frac{1.63 \times 10^{-26} \mbox{cm}^3/\mbox{s}}{\langle \sigma v \rangle } \right) \, .
\ee
where we have used GeV$^{-2} \cdot c \simeq 1.17 \times 10^{-17}$~cm$^3/$s.  WIMPs with typical speeds ($v\simeq 0.3 c$) and electroweak cross-sections ($\approx 1$~pb) yield  $\Omega_{dm}^{therm}h^2\simeq 0.12$ in  agreement with the data, a  coincidence often called  the {\em WIMP miracle}. 

\begin{figure}[t]
\begin{center}
\includegraphics*[width=4.0in]{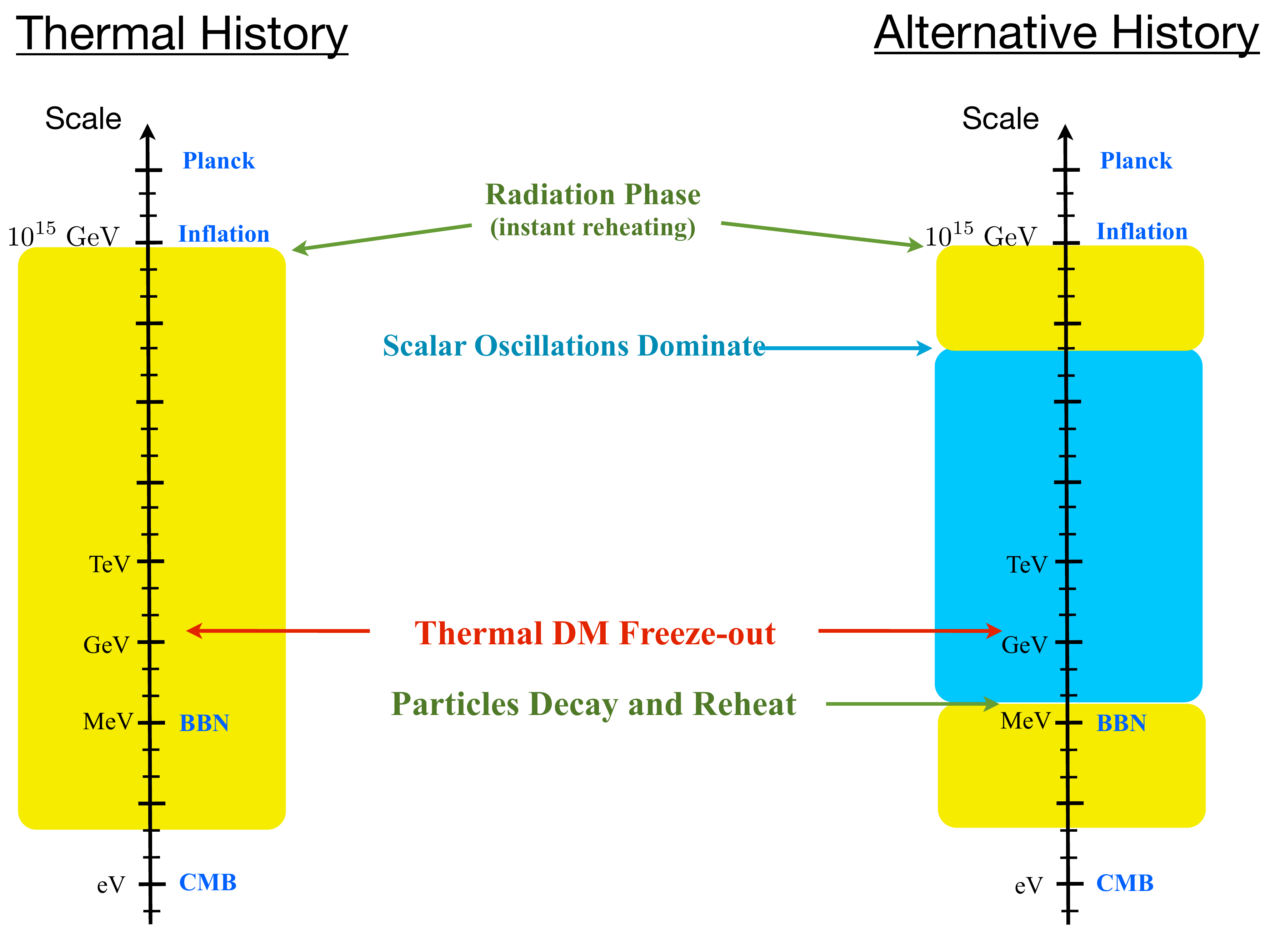}\label{fig1}
\hspace{0.2cm}
\end{center}
\caption{The lefthand timeline represents the thermal history of the early universe when dark matter is populated in the thermal bath that emerges shortly after after inflation.  The right timeline represents a possible nonthermal history where dark matter production occurs directly from scalar decay.}
\end{figure}

Simple SUSY models with thermal WIMPs are in growing conflict with collider data and  direct detection experiments \cite{Baer:2011ab}. By contrast, nonthermal models posit that  dark matter production occurs at temperatures below standard thermal freeze-out\footnote{If the particles were produced above their freeze-out threshold, they could thermalize via their mutual interactions.}  leading to dark matter with novel and unexpected experimental signatures.  For example, if a heavy relic comes to dominate the energy density following inflation and the dark matter particle is one its decay products, the resulting relic density is still given by (\ref{totalthermaldm}) but  with $T=T_r$ and $g_*=g_*(T_r)$, the value at the time of reheating
\bea \label{nonthermal_dm}
\Omega_{dm}^{NT}h^2 &\simeq& 8.60 \times 10^{-11} \left(
\frac{m_X }{g_*(T_r)^{1/2} \langle \sigma v \rangle T_r} \right) \, , \nonumber \\
&\simeq& 0.10  \; \left( \frac{m_X}{100~\mbox{GeV}} \right) \left( \frac{10.75}{g_*} \right)^{1/2} \left(
\frac{3 \times 10^{-23} \, \mbox{cm}^3/\mbox{s}}{\langle \sigma v \rangle} \right)\left( \frac{10 ~\mbox{MeV}}{T_r} \right) \, .  
\eea
The similarity to the thermal freezeout result (\ref{totalthermaldm}) arises because when the WIMPs are produced from scalar decay they will  
rapidly annihilate until their number density reduces to the point where annihilations can no longer occur.  
This process is essentially instantaneous (on cosmological time scales) and so this second ``freezeout'' occurs at the reheat temperature $T_r$ (see \cite{Watson:2009hw} for a review). 
 Any thermally produced dark matter is diluted by the increase in entropy during the decay by a factor of $\left( T_r / T_f \right)^3$.  Equation~(\ref{nonthermal_dm}) demonstrates both the benefits and disadvantages of  nonthermal dark matter.  There is no longer a robust relationship between $m_X$ and and freeze-out temperature but  there is  more flexibility  to satisfy the observational constraint $\Omega_{dm}h^2=0.12$, and the possibility of larger annihilation rates. The extreme case of MeV scale reheating enhances the annihilation rate by three orders of magnitude, relative to the thermal WIMP case.  This would yield larger  fluxes in indirect detection experiments, and modifies design  strategies for direct detection and collider probes. This has led to new model building possibilities for SUSY neutralino dark matter, many of which are already tightly constrained by PAMELA and FERMI \cite{Grajek:2008pg,Grajek:2008jb,Dutta:2009uf,Kane:2009if,Sandick:2011rp}.   

There have been several phenomenological studies of nonthermal dark matter over the years. This option became more attractive when it was realized that in SUSY based solutions to the hierarchy problem -- where gravity is important -- the reheat temperature is not a free parameter, but is fixed by the high energy behavior of the theory  \cite{Acharya:2008bk,Acharya:2009zt}.   In combination with tightening collider and dark matter detection constraints on thermal dark matter there is thus considerable motivation for considering nonthermal dark matter.

A comparison of the thermal history of the universe for representative thermal and nonthermal scenarios appears in Figure 1. There are many  possible alternatives to a strictly thermal history, which are generally associated with   dark matter production that occurs at a temperature below that of thermal production, or out of equilibrium.  These can include cosmic histories where there is a second phase of low-scale inflation (thermal inflation \cite{Lyth:1995ka}), or if the decay of heavy particles leads to a significant source of dark matter and entropy production prior to BBN.

\subsection{Nonthermal Dark Matter:  A Realization Through Scalar Decay \label{mot*}}

Many Beyond the Standard Model (BSM) proposals contain scalar degrees of freedom beyond the minimal higgs. This is the case in supergravity and string theoretic approaches to BSM, where the vacuum expectation values of scalar fields determine the couplings of the low energy theory.   However, these fields also lead to the cosmological moduli problem \cite{Coughlan:1983ci,deCarlos:1993jw,Banks:1993en} -- the fields are displaced from the low-energy minima in the early universe and undergo coherent oscillations, mimicking a matter dominated epoch\footnote{This is strictly true only if the mass term in the potential gives the dominant contribution, otherwise the cosmological scaling of the energy density is determined by the dominant term in the potential~\cite{Turner:1983he}.} prior to BBN.
The fields typically decay through gravitational strength couplings, and the universe reheats via the production of relativistic Standard Model  and BSM particles -- the lightest of which, if stable, may provide a WIMP candidate.

For scalars of mass $m_\sgm$ the decay rate typically scales as $\Gamma \sim m_\sgm^3 / m_p^2$
and the corresponding reheat temperature is
\be
T_r \simeq \left( \frac{m_\sgm}{10 \; \mbox{TeV}} \right)^{3/2} \; \mbox{MeV}.
\ee
If this  temperature is below the thermal freeze-out scale $T_f \simeq m_X/20$
 and the field dominates the energy density at the time of decay we have a nonthermal  dark matter scenario. Successful BBN and observations of neutrino decoupling require $T_r \gtrsim 3~$MeV~\cite{Hannestad:2004px,Kawasaki:2000en,Kawasaki:1999na,DeBernardis:2008zz}. For dark matter with a mass not too far above the electroweak scale, fixing $T_r < T_f \sim m_x / 20$ provides an upper bound 
\be \label{therange}
 20 \; \mbox{TeV}  \lesssim m_\sgm \lesssim 10^4 \; \mbox{TeV}.
\ee

To give a  specific example\footnote{The arguments that follow will not rely strongly on the presence of SUSY, and it would seem that the main ingredients in our argument -- the existence of scalars and symmetry breaking associated with the electroweak scale -- could be realized in other approaches to BSM physics.}, consider a supersymmetric model with a singlet scalar field $\sgm$ with a shift symmetry $\sgm\to\sgm + c $ where $c$ is a constant, so the potential is independent of the field, or $V(\sgm)=0$. If this remains a good symmetry until SUSY breaking and SUSY breaking is mediated by gravitational interactions the resulting mass is comparable that of the gravitino $m_{3/2}$. If SUSY  addresses the electroweak hierarchy problem 
\beq
m_{3/2}=\fr{\Lambda^2}{m_p}\simeq 0.1 - 10^{3} ~ \mbox{TeV},
\eeq
where $\Lambda$ is the SUSY breaking scale, the electroweak hierarchy implies $\Lambda\simeq 10^{11}-10^{12}$~GeV.  Thus,  SUSY theories can easily lead to masses in the range given by equation~(\ref{therange}).  The lower bound of $0.1$~TeV is the origin of the term {\em cosmological moduli problem}, as it leads to scalar decay and a  reheat temperature in conflict with the bounds set by BBN and neutrino decoupling. Within the MSSM, such low scales are already disfavored by LHC data since squarks in this mass range have not been detected, pushing up the mass scale of the gravitino.  

A shift-symmetric scalar in a fundamental theory -- such as supergravity or string theory --  typically has additional geometric factors that lift its mass to larger values.  For Type-IIB flux compactifications   the mass is of order $m_\sigma \sim \log(m_p / m_{3/2}) m_{3/2}$ \cite{LoaizaBrito:2005fa} if the model accounts for both the electroweak hierarchy and the present-day vacuum energy.  The authors of \cite{Acharya:2009zt}  argued that in supergravity and string frameworks the mass of a scalar which is stabilized and which meets the above requirements is typically within the range of equation~(\ref{therange}), implying a nonthermal history. 

In the early universe and during inflation,  the shift symmetry is broken by both the finite energy density of the universe and quantum gravity effects, contributing a Hubble scale mass and a tower of non-renormalizable operators to the effective potential,
\beq
\Delta V(\sgm) = -c_1 H^{2}_{inf} \sgm^{2}+ \fr{c_2}{m^2_p} \sgm^{6}+ \dots ~ ,
\eeq
where we expect the couplings $c_1,c_2 \simeq \mathcal{O}(1)$. During high scale inflation $H>m_\sgm$ and  $\langle \sgm \rangle \simeq m_p$, as opposed to the low-energy minimum $\langle \sgm \rangle \simeq 0$ resulting from SUSY breaking.  The displacement from the low energy minima provides the initial amplitude for the coherently oscillating field $\sgm$. The energy density of the coherent field is  
\beq\label{rho_osc}
\rho_{osc}^{\sgm}(t)= \fr{1}{2}m^2_{\sgm}\Delta\sgm^2 \left(\fr{a(t_{osc})}{a(t)}\right)^3.
\eeq
Coherent oscillations begin as the expansion rate reaches $H \simeq m_\sgm$, corresponding to a temperature 
\beq\label{T_osc}
T_{osc}= \left(\fr{\pi^2g_*(T_{osc})}{90}\right)^{-1/4} (m_\sgm m_p)^{1/2}\simeq 2.25\times 10^{11} \left(\fr{g_*(T_{osc})}{200}\right)^{-1/4} \left(\fr{m_\sgm}{100~TeV}\right)^{1/2}~\mbox{GeV}.
\eeq
 The Universe remains effectively matter dominated until the field decays into Standard Model and SUSY particles when $\Gamma_{\sgm}\simeq H$. For a gravity mediated process, the  decay rate is 
\beq\label{d_rate}
\Gamma_{\sgm}= c_3 \fr{m^3_{\sgm}}{m^2_p},
\eeq
where $c_3 = 1/(4 \pi)$ is a typical value. 
At the time of decay the transfer of energy from the scalar field to Standard Model and SUSY particles will be instantaneous compared to the expansion rate and, because the scalar dominates the energy density, we expect a large yield of dark matter and radiation\footnote{If the decay to SUSY particles was for some reason further suppressed compared to the Standard Model, amount of dark matter would be set by the corresponding branching ratio and the initial amount of scalar field condensate.  Such a situation is difficult to arrange in practice.}.  The radiation represents the relativistic Standard Model particles, whereas the dark matter results from rapid decays of  SUSY particles down to the Lightest SUSY Particle (LSP).  Due to the large production of LSPs, some annihilations take place, and these particles will achieve kinetic equilibrium quickly by scattering off the relativistic bath of Standard Model particles.

The  reheat temperature of the universe is
\beq\label{T_r}
T_r=\left(\fr{\pi^2g_*}{90}\right)^{-1/4}(\Gamma_\sgm m_p)^{1/2} \simeq 20~c_3^{1/2} \left(\fr{g_*}{10.75}\right)^{-1/4}\left(\fr{m_\sgm}{100~\mbox{TeV}}\right)^{3/2}~\mbox{MeV}.
\eeq
and $g_*\equiv g_*(T_r)=10.75$ if $T_r$ is low. Using this expression and (\ref{nonthermal_dm}), we  estimate the relic density in nonthermal dark matter, 
\bea  \label{nt1}
\Omega_{dm}^{NT}h^2 &\simeq& 8.60 \times 10^{-11} \left(
\frac{m_X }{g^{1/2}_* \langle \sigma v \rangle T_r} \right) \, , \nonumber \\
&\simeq& 0.08 \left( \frac{m_X }{g_*^{1/4} \langle \sigma v \rangle m_\sigma^{3/2}} \right),
\eea
which now depends only on the properties of the dark matter (mass and annihilation rate) and the mass of the decaying scalar resulting from SUSY breaking. As discussed above, in most models the scalar mass is not a free parameter, but similar to the gravitino mass $m_{3/2}$, which is related to the scale of SUSY breaking as $\Lambda^2_{susy}=m_{3/2}m_p$.  Thus, the mass of the scalar (and so the relic density of dark matter) is controlled by the need for SUSY to generate a hierarchy between the electroweak and Planck scale (i.e. $\Lambda_{EW} \sim m_{3/2} \ll m_p$).  With a typical SUSY breaking scale of $\Lambda = 10^{11}$~GeV, corresponding to a gravitino mass of around $4$~TeV the resulting relic density is
\bea  
\Omega_{dm}^{NT}h^2  &\simeq&0.11 \; \left( \frac{m_X}{100\gev} \right)\left( \frac{10.75}{g_*} \right)^{1/4}\left( \frac{3 \times 10^{-23}\cms}{\langle \sigma v \rangle} \right)\left( \frac{4~\mbox{TeV}}{m_{3/2}} \right)^{3/2} \left( \frac{34}{k} \right)^{3/2}
\eea
where we have set $c_3=1/(4\pi)$, and the ratio between the scalar and gravitino mass as
 $k=m_\sigma / m_{3/2} \simeq \log( m_p / m_{3/2} )$ -- which is only logarithmically sensitive to changes in the hierarchy.  This constant is model dependent and typically between ${\cal O}(1-100)$.  
We have chosen a fiducial value for the annihilation rate that yields roughly the right amount of dark matter  for the  hierarchy set by the choice of low-scale SUSY breaking $\Lambda=10^{11}\gev$.  The  cross-section is three orders of magnitude higher than expected with a thermal history with important experimental consequences, as discussed in Section 4.

The reheat temperature in this framework is not a free parameter,  
but  a consequence of the hierarchy between the electroweak and Planck scale (determined by $\Lambda_{susy}^2=m_{3/2} m_p$), which also helps determine the SUSY breaking masses of other sparticles in the theory. In both supergravity and string motivated approaches, the key lesson is that the reheat temperature is intimately connected to other aspects of the theory and not a free and tunable parameter.  Given that gravitationally coupled scalars are generic in  high energy completions of the Standard Model and in no sense exotic, we see that nonthermal histories are a feasible and robust possibility.

\subsection{Nonthermal Histories and CMB Observables} \label{moduli_decay}

\begin{figure}[p!]
\begin{center}
\includegraphics[scale=1.2]{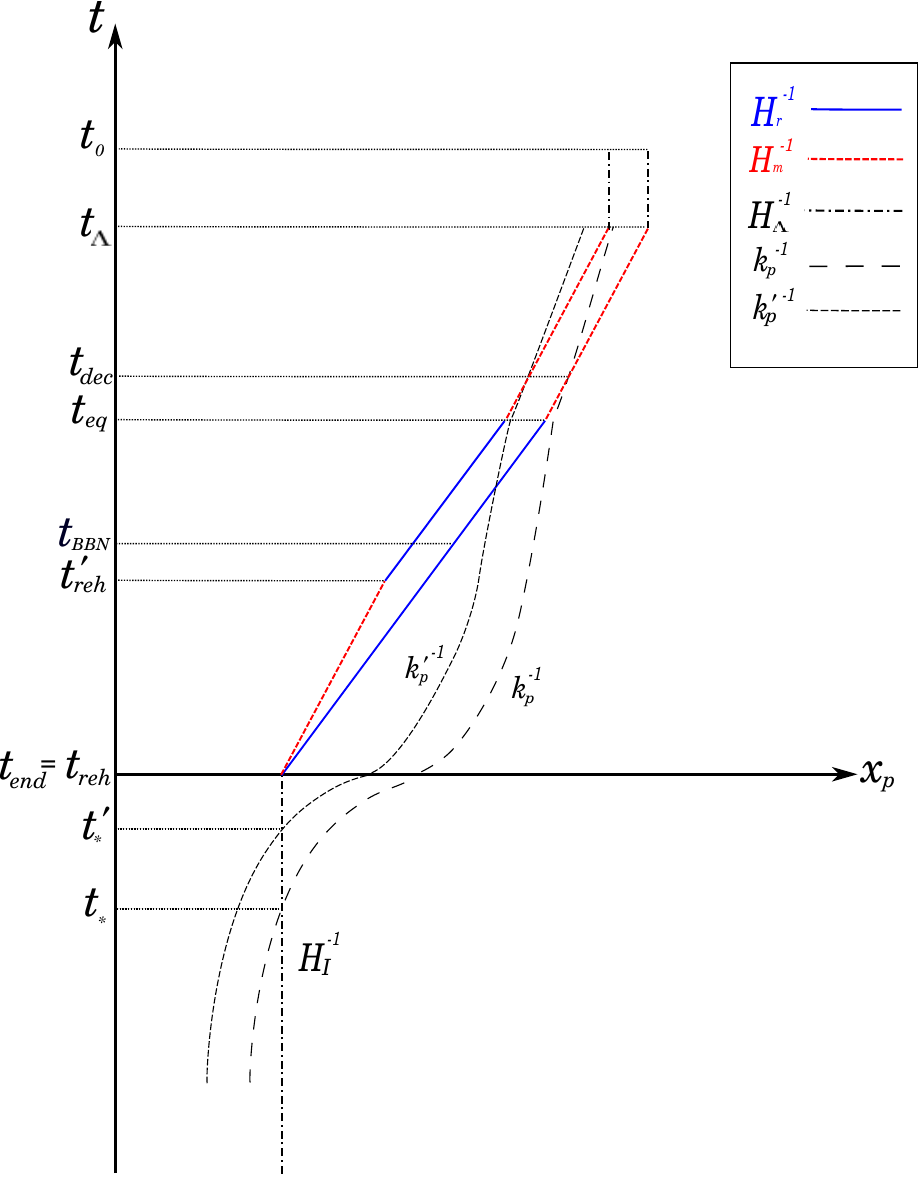}
\hspace{0.5cm}
\end{center}
\caption{Evolution of physical wavelengths as labelled by their inverse wavenumber $k_p^{-1}$ during inflation (below the x-axis) and during the post-inflationary epoch (above the x-axis). The solid (blue) line represents the Hubble radius, $\textcolor{blue}{H_r^{-1}}$ in a Universe dominated by a radiation fluid $w=1/3$, the dashed (red) line is the Hubble radius, $\textcolor{red}{H_m^{-1}}$ in a post-inflationary era dominated by a pressure-less fluid, $w=0$. We compare the evolution of a physical mode $k_*$  that re-enters at CMB decoupling in the standard scenario (Radiation $\rightarrow$ Matter $\rightarrow$ Dark energy) with a mode $k^{'}_{*}$ that re-enters at CMB decoupling in the nonthermal scenario (Matter$\rightarrow$ Radiation$\rightarrow$ Matter$\rightarrow$ Dark Energy).
These modes exit the Hubble radius at different times during inflation,  $t_*$ and $t^{'}_*$, which translates into a shift in the number of e-folds $\Delta N= H \Delta t$. The corresponding shift in the pivot scale or any co-moving mode is given by $k_*'=k_* e^{-\Delta N}$.}\label{fig:spacetime}
\end{figure}

 For simplicity we  assume inflationary (p)reheating was instantaneous (on gravitational time scales) and focus on the oscillations of the scalars, which come to dominate the energy density, as specified by equation~(\ref{T_osc}).  
Following \cite{Liddle:2003as,Kinney:2005in,Adshead:2010mc}, we make the substitution $\rho_{end}\rightarrow\rho_{osc}^\sigma$, using $\rho_{r}^\sigma=(\pi^2/30) g_*(T_r^\sigma)(T_r^{\sigma})^4$. At the onset of oscillations $\rho_{osc}^{\sgm}(t_{osc})= \fr{1}{2}m^2_{\sgm}\Delta\sgm^2$ and
\bea
\Delta N &=& -0.04 + \frac{1}{12}\ln \left( \frac{g_*(T_r^\sigma) T_r^4}{m_\sigma^2 \Delta\sigma^2} \right), \nonumber \\
&=&-10.75 + \frac{1}{12}\ln \left[       \left(\frac{g_*(T_r^\sigma)}{10.75} \right) \left( \frac{T_r}{3 ~\mbox{MeV}} \right)^4 
 \left(\frac{100 ~\mbox{TeV}}{m_\sigma} \right)^2 \left( \frac{m_p}{\Delta \sigma} \right)^2 \right], \label{thedeltaN} 
\eea
where we used $w=0$, and the second line expresses the parameters relative to fiducial values.  If  scalar decay proceeds via a gravitational strength coupling, equation~(\ref{d_rate}) eliminates the mass dependence in (\ref{thedeltaN}). With $c_3=1/(4\pi)$ we find
\be \label{deltaNred}
\Delta N=-10.68 + \frac{1}{18}\ln \left[       \left(\frac{g_*(T_r^\sigma)}{10.75} \right) \left( \frac{T_r}{3 ~\mbox{MeV}} \right)^4 
  \left( \frac{m_p}{\Delta \sigma} \right)^3 \right].
\ee 
This shift and its effect on physical modes is described qualitatively in Figure \ref{fig:spacetime}. We see that $\Delta N$ is  logarithmically sensitive to changes in  parameters, including the reheat temperature.  To generate nonthermal dark matter, the reheat temperature must typically be below about $T_r\simeq 10$ GeV, but above the BBN and neutrino bounds of about $T_r\simeq 3$ MeV.  The range of possible temperatures is more than four orders of magnitude, but (\ref{deltaNred}) the corresponding shift in $\Delta N$ is  $-10.68\lesssim \Delta N \lesssim -8.51$.  Thus, for the scenarios considered here we have a relatively robust   $|\Delta N | \simeq 10$.   Physically, a more massive field decays earlier while a lighter field decays later, but oscillations and the corresponding matter dominated phase also begin later as well (as seen from \eqref{T_osc}), leading to similar values of $\Delta N$.     

The change in inflationary observables is estimated by recalling that for most simple models of inflation, the running of the spectral index $\alpha_s \equiv d n_s /d \log{k}$ is  typically  $-10^{-4} \gtrsim \alpha \gtrsim   -10^{-3}$ \cite{Adshead:2010mc}, so $\Delta n_s$ is between $-10^{-3}$ and $-10^{-2}$, relative to the value seen with instant reheating.  The remaining uncertainty in the $n_s$ and $r$ is significantly  reduced, since these models  predict that the universe is matter dominated through most of the primordial dark age. 

\section{Constraining Nonthermal Dark Matter}

   We focus on SUSY neutralinos as the WIMPs,  but we expect our conclusions to be easily extended to other non-SUSY dark matter candidates.  The neutralino is an electrically charge neutral state and linear combination of the superpartners of the Standard Model $B$, $W^3$, and higgses\footnote{The MSSM extension of the Standard Model higgs sector requires two higgs doublets.} 
\be \label{compose}
\chi^0=N_{10} \tilde{B} + N_{20} \tilde{W}^3 + N_{30} \tilde{H}_1^0+N_{40}\tilde{H}_2^0,
\ee
where $\tilde{B}$ and $\tilde{W}^3$ are the bino and wino, and $\tilde{H}_{1,2}$ are higgsinos.  The $N_{i0}$'s  denote the amount each component  contributes to the neutralino.  The neutralino or WIMP mass is determined by\footnote{We refer the reader to \cite{Jungman:1995df} for more details.} diagonalizing a  matrix which depends on the masses of the bino, wino, and higgsino ($M_1$, $M_2$, and $\mu$, respectively), the Weinberg angle $\theta_W$, and $\tan(\beta)$ which is the ratio of the vacuum expectation values of the higgs vevs.

When dark matter is composed of thermally produced neutralinos, the neutralino must be bino-dominated, which causes  neutralinos to annihilate less efficiently, generating the correct relic density of dark matter \cite{Jungman:1995df}. However, if the reheat temperature following scalar decay is below thermal freeze-out, larger annihilation cross-sections are required.
Likewise, because the decaying scalar gets a mass from SUSY breaking there is a 
natural relationship between the reheat temperature and the scale of SUSY breaking, which addresses the hierarchy problem and sets other sparticle masses.  The hierarchy problem requires $m_\sigma \sim m_{3/2} \sim$~TeV, which results in reheat temperatures below neutralinio freeze-out temperature ($T_f \simeq m_X/20$),  favoring a nonthermal history.  The resulting dark matter density is given by (\ref{nonthermal_dm}) as
\bea \label{ntsw}
\Omega_{dm}^{NT}h^2 &\simeq& 0.10  \; \left( \frac{m_X}{100~\mbox{GeV}} \right) \left( \frac{10.75}{g_*} \right)^{1/2} \left(
\frac{3 \times 10^{-23} \, \mbox{cm}^3/\mbox{s}}{\langle \sigma v \rangle} \right)\left( \frac{10 ~\mbox{MeV}}{T_r} \right) \, ,  
\eea
requiring a larger annihilation cross-section. For neutralinos, the larger cross-section requires a more significant contribution from wino and higgsinos, changing expectations for colliders, and for direct and indirect detection experiments.  

Existing data from these experiments place a lower bound on the reheat temperature.
From (\ref{ntsw}), with Planck's central value $\Omega_{dm}h^2 \simeq 0.12$ and constraints from indirect detection  on $\sigma v$ and solving for the reheat temperature we  arrive at a minimum value which is typically above the hard lower bound of  $3$~MeV, further constraining the equation of state in the primordial dark ages. 

\subsection{Nonthermal Wino-like Neutralinos} 
\begin{figure}[t!] 
\includegraphics[scale=0.58]{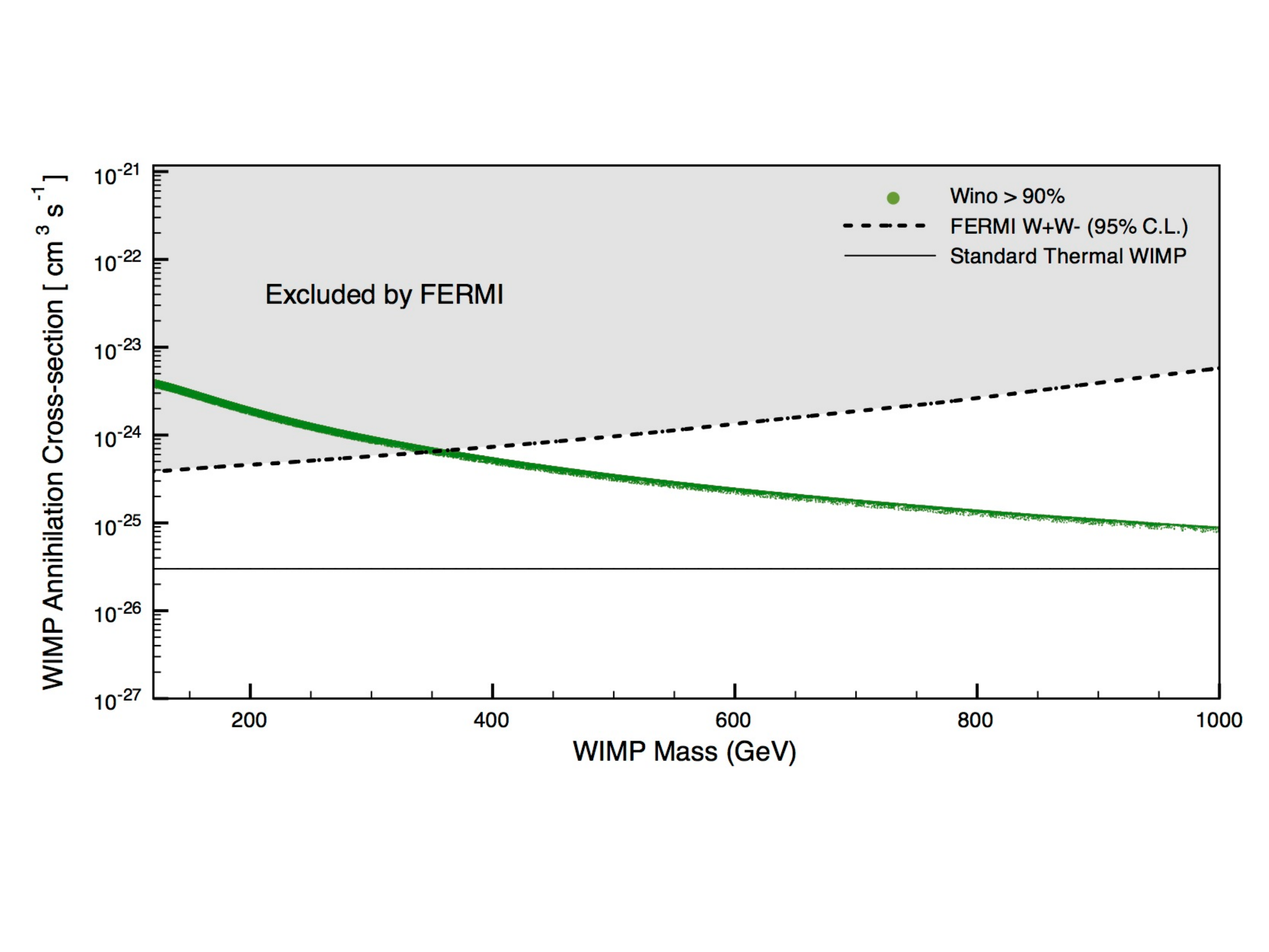}
\vspace{-0.9in}
\caption{ \label{winofig}  The thermally average annihilation rate $\langle \sigma v \rangle$ for a dominantly wino neutralino to annihilate to a pair of $W$-bosons, as a function of mass. The Fermi constraint comes from two years of data from 10 Dwarf spheroidal galaxies \cite{Ackermann:2011wa}.     These results have been obtained using  DarkSUSY  \cite{Gondolo:2004sc}, but the general shape of the curve is in good agreement with the analytic expression (\ref{winoann}).
For this scan we took the MSSM parameters to vary over: $M_2=100$~GeV to $2$~TeV, $\mu=100$~GeV to $2$~TeV, and $\tan \beta=5$ to $50$.  We  applied all LEP2 constraints and color charged particles were taken to decouple by setting their masses to be above $2$~TeV, allowing agreement with LHC constraints.}
\end{figure}
The thermally averaged cross-section for the dominantly wino-like neutralino is given by~\cite{Moroi:1999zb}
\be \label{winoann}
\langle \sigma v \rangle = \frac{g_2^4}{2\pi M_2^2} \left( \frac{(1-x_W)^{3/2}}{(2-x_W)^2} \right) 
\ee 
where $x_W \equiv m_W^2/M_2^2$ ($m_W$ is the mass of the W-boson), $g_2 \simeq 0.66$ is the $SU_L(2)$ electro-weak gauge coupling (at the weak scale) in the MSSM, $M_2$ is the wino mass, and we note that the result is independent of the velocity (s-wave channel). From (\ref{winoann}), a wino of mass $M_2 = 100$~GeV, the annihilation rate will be around $\langle \sigma v \rangle = 4.06 \times 10^{-24} \cms$, exceeding the cross-section expected for thermal WIMPs by about two orders of magnitude.  The cosmological constraint (\ref{ntsw}) then requires a reheat temperature of around $67$~MeV.  From (\ref{winoann}) we see that as the wino mass increases, the corresponding annihilation rate  decreases, requiring larger reheat temperatures via (\ref{ntsw}). At this point it seems that the reheat temperature is a free and tunable parameter.  However, additional experimental constraints can be placed on the wino cross-section through the indirect detection of dark matter.

The wino annihilation rate is s-wave\footnote{We note that annihilations with other light MSSM states (coannihilations) can be crucial when calculating the relic density \cite{ArkaniHamed:2006mb}, and for high mass winos ($m_X>>$~TeV) Sommerfield enhancement may also play an important role \cite{Moroi:2013sla}. However, for the range of masses and temperatures we will consider (in order to establish a {\em lower bound} on the reheat temperature) these effects are negligible.} so the annihilation rate above remains relevant for winos in the galaxy today, which are non-relativistic with $v \simeq 10^{-3}c$.  Annihilation is dominantly into W-boson pairs,  providing a source of anti-protons, positrons, and gamma rays.  Indirect detection measurements constrain the cross-section, but suffer from a number of astrophysical complications, which includes uncertainties in the halo profile and propagation models \cite{Kane:2011zz}).  Therefore,  the best constraints on the wino arguably come from gamma rays as opposed to charged anti-matter, and we will use bounds from FERMI's two year data from observations of 10 Dwarf Spheroidal galaxies \cite{Ackermann:2011wa}, showing our results for the cross-section in  Figure \ref{winofig}.  For masses less than roughly $375$~GeV the wino annihilation rate is too large, giving $\langle \sigma v \rangle \lesssim 6.13 \times 10^{-25} \cms$.  Using this in the cosmological constraint (\ref{ntsw}) we find $T_r \gtrsim 696~\mbox{MeV}$ (where $g_\ast = 61.75$). Finally, the corresponding change in the number of e-folds from (\ref{deltaNred}) is $\Delta N=-9.37$. 

\begin{figure}[t!]
\includegraphics[scale=0.43]{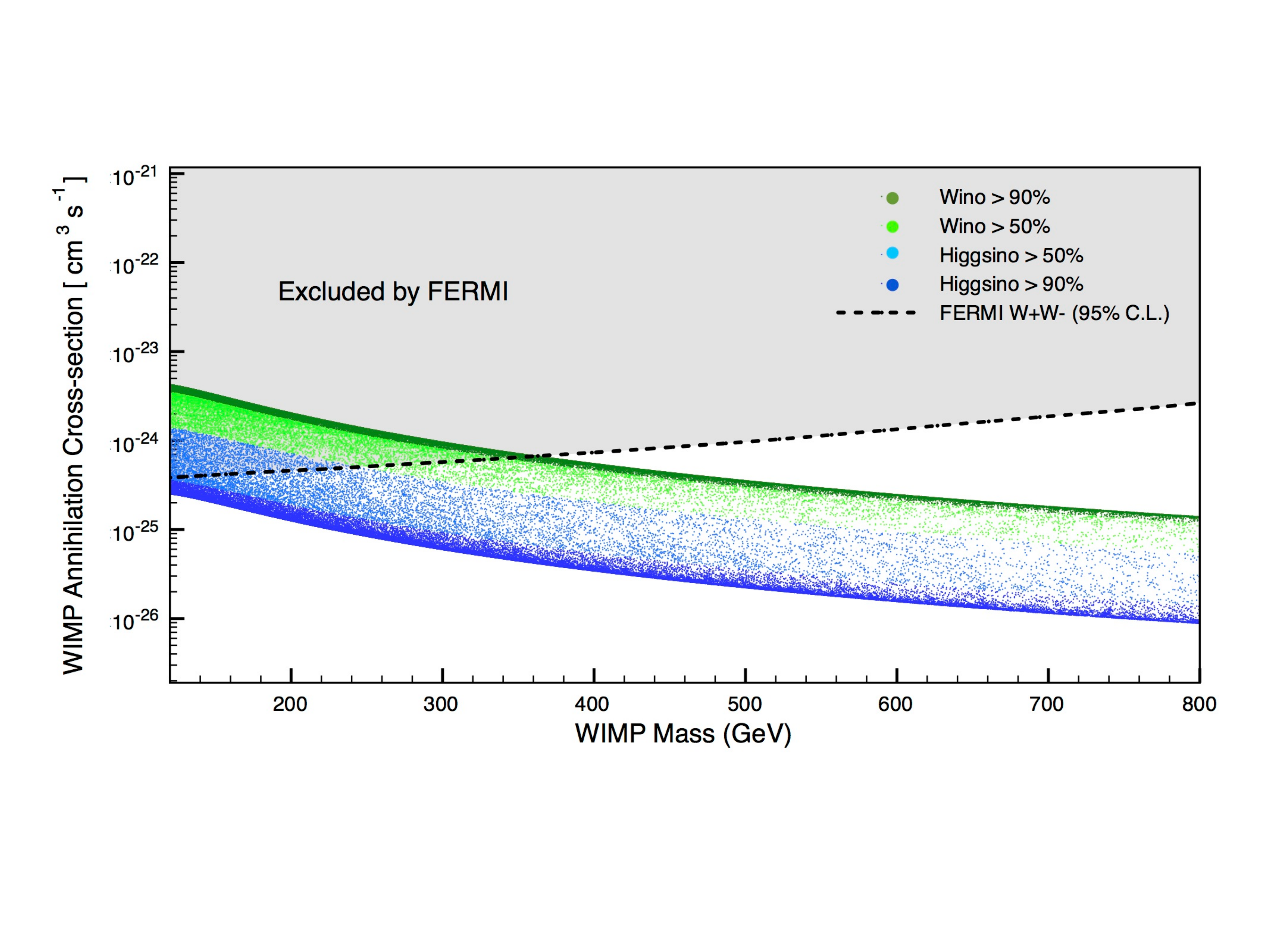}
\vspace{-0.9in}
\caption{\label{fig:indirectall} The thermally average annihilation rate $\langle \sigma v \rangle$ for a general neutralino to annihilate to a pair of $W$-bosons, with a bino fraction of less than $10\%$, to realize a nonthermal history.  The constraint from Fermi comes from two years of data from 10 Dwarf spheroidal galaxies \cite{Ackermann:2011wa}.  These results have been obtained using  DarkSUSY  \cite{Gondolo:2004sc}, however the general shape of the upper curve is in good agreement with the analytic expression (\ref{winoann}) and the shape of the lower, higgsino curve agrees with the expectation that $\langle \sigma v \rangle \sim 1/\mu^2$. Other parameter choices match those in Figure~\ref{winofig}. }
\end{figure}
\begin{center}
\begin{figure}[t!]
\includegraphics[scale=0.58]{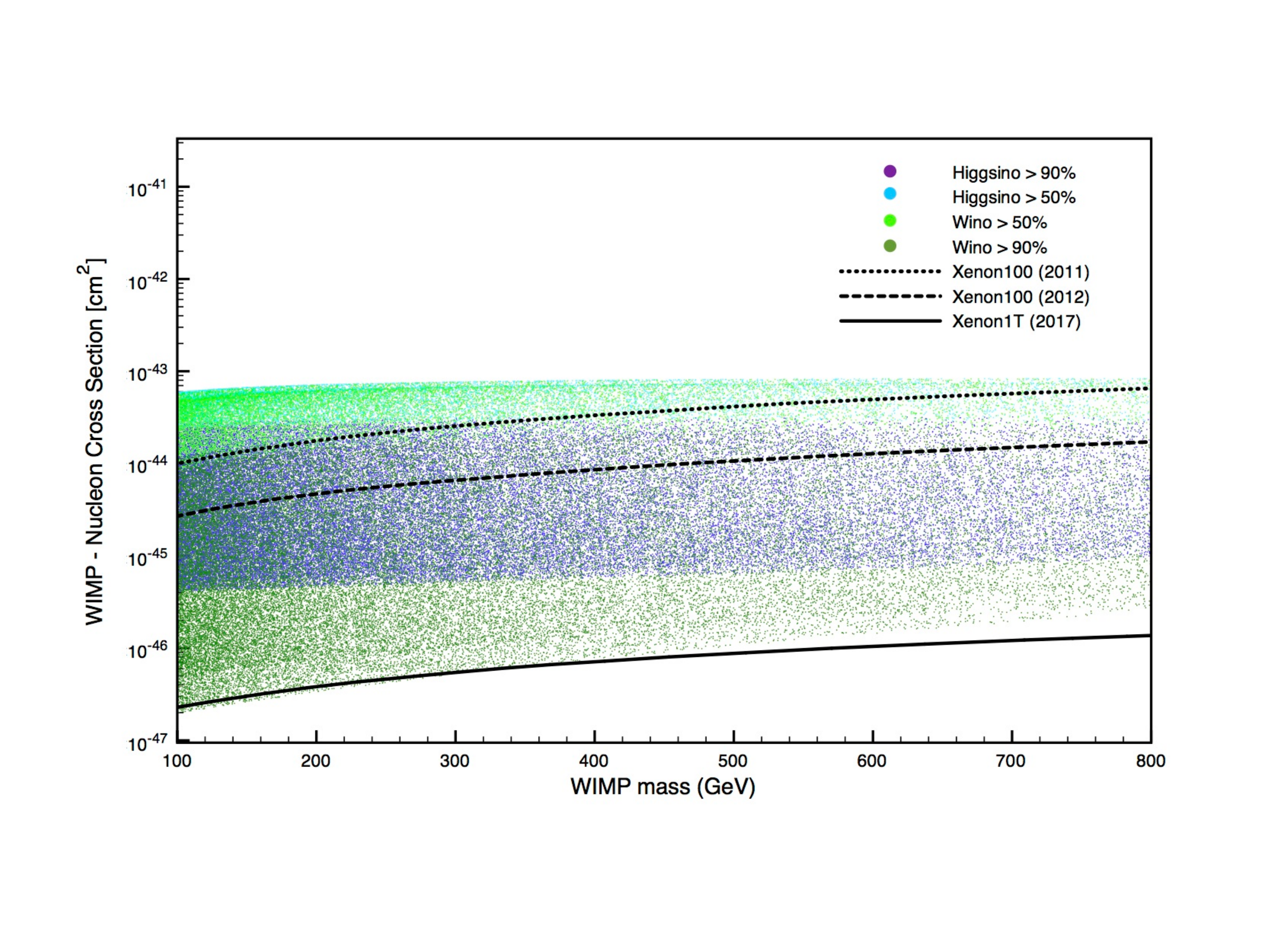}
\vspace{-0.5in}
\caption{\label{fig:dd} 
The WIMP-nucleon (proton) scattering cross-section as a function of WIMP mass. For wino-higgsino mixtures we find that most models are excluded by the Xenon 2011 / 2012 data.  For purified WIMPs (dominantly wino or higgsino) many models escape existing constraints and for models with wino fractions $90\%$ we must wait until Xenon1T for meaningful constraints to be established.   However, for the dominantly higgsino models many are already disfavored.  For this scan we took the MSSM parameters to vary over: $M_2=100$~GeV to $2$~TeV, $\mu=100$~GeV to $2$~TeV, and $\tan \beta=5$ to $50$.  We have applied all LEP2 constraints and color charged particles were taken to decouple by setting their masses to be above $2$~TeV -- allowing agreement with LHC constraints.}
\end{figure}
\end{center}
\subsection{Neutralino WIMPs:  The General Case}
Neutralinos can also contain bino and higgsinos in their composition as indicated in (\ref{compose}). We now consider  more general  neutralinos within the nonthermal framework.  
For a large bino contribution to (\ref{compose}) the annihilation rate is  too small to allow a nonthermal history. 
Thus,  we  restrict the bino fraction to be less than $10\%$  to ensure that a nonthermal history is realized\footnote{We refer the reader to  \cite{Feldman:2013gfa} for a recent account of the phenomenology of bino-mixed neutralinos as thermal dark matter and their observational consequences.}.
On the other hard, a neutralino with a large higgsino component is compatible with a nonthermal history.  In Figure \ref{fig:indirectall} we  present the FERMI constraints on annihilations to W-bosons allowing for this possibility.  We scan the MSSM parameter space using DarkSUSY \cite{Gondolo:2004sc} and present results for around $100,000$ models.  
We restrict the bino-fraction to be less than $10\%$, and we take $\mu$ and the wino mass ($M_2$) to range from $100$~GeV up to $2$~TeV, and $\tan \beta$ between $10-50$. 

We reject models that are incompatible with collider data, but  properties of neutralino WIMPs are primarily determined by the gaugino masses ($M_1$ and $M_2$), $\mu$ and $\tan \beta$ -- see e.g. the recent discussion in \cite{Cheung:2012qy}.  Thus, it is  easy to obtain models consistent with LHC constraints on color charged super-partners. We also require a $126$~GeV higgs\footnote{There are constraints from LHC on light neutralinos, but because we are considering masses larger than around $100$~GeV these constraints are not important here \cite{ATLAS:2012uks,Chatrchyan:2012pka}.}.  From Figure \ref{fig:indirectall}, we see that our numeric results agree well with the analytic expectation that a pure wino annihilation rate should scale as $1/M_2^2$ (top curve in Figure \ref{fig:indirectall}), whereas a pure higgsino would  scale as $1/\mu^2$ (bottom curve in Figure \ref{fig:indirectall}).  Allowing for a higgsino contribution relaxes the bound on the reheat temperature provided by FERMI -- with  a pure higgsino being completely unconstrained.  

We have restricted  attention to the $W$-boson annihilation channel, which is typically dominant for well-mixed neutralinos, but we find similar  constraints for annihilations to other common channels such as bottom quarks.  Our key observation is that indirect detection alone does  not put a useful bound on the reheat temperature when more general neutralinos are considered.  However, bounds from direct searches partially remedy this situation. Recall that a pure wino-like neutralino gives little direct detection signal, as the WIMP-nucleon interaction is loop suppressed \cite{Cheung:2012qy} but for more general neutralinos the situation changes and direct detection experiments provide meaningful constraints. 
 
Consider the spin independent constraints provided by  Xenon100  \cite{Aprile:2012nq}, 
as well as  future constraints expected from Xenon1T \cite{Beltrame:2013bba}.
In Figure \ref{fig:dd} we use the Xenon 2011 and 2012 null results to constrain the nonthermal neutralino models considered above. Although  higgsino mixing  relaxed  the constraint on the reheat temperature coming from FERMI, many of these models are then ruled out by Xenon100.  As seen in Figure~\ref{fig:dd},  unless the neutralino is purely wino or higgsino,  it is typically in tension with the Xenon100 data.   Xenon1T will constrain these models even further, and would potentially bringing the pure wino into tension if it yields a null result. 
Thus, for generically mixed neutralinos, the nonthermal history is in tension with direct detection data, and for the pure wino  the lower bound on the reheat temperature is  $696$~MeV.  The only exception is the pure higgsino, which in the low mass range is somewhat constrained by direct detection but does allow in some cases for a lower reheat temperature.

\section{Conclusion}
Current LHC constraints on scalar super-partner masses suggest a new mass scale $m_{3/2}=\Lambda^2/m_p$ around the $10-100$ TeV range. When the MSSM is accompanied by additional singlets which receive SUSY breaking masses near this scale, this implies a nonthermal history for the early universe. We have shown that a nonthermal history modifies the predictions of inflationary models relative to those seen with a thermal history, and that these changes are comparable to the precision of parameter estimates made with Planck data.

A caveat to our analysis is provided by the recent work in \cite{Cheung:2012qy} (see also \cite{ArkaniHamed:2006mb,Hooper:2013qjx}),  showing that there are certain 
regions of the neutralino parameter space `hidden' to direct detection experiments.  Although one may expect such points to be atypical, it has been argued
for some time that special relations between parameters (e.g. in the case of well-tempered neutralinos \cite{ArkaniHamed:2006mb}) may be the only way for SUSY based WIMPs to survive, given existing collider constraints. We leave a more detailed analysis -- including these subtleties and constraints from spin-dependent interactions -- to future work.   In addition, perturbations grow in a matter dominated universe, so  density inhomogeneities with an initial amplitude of $\delta \rho /\rho \sim 10^{-5}$  grow to be of order unity during the matter dominated phase, a phenomenon also seen in inflationary models with inefficient reheating \cite{Easther:2010mr}. Consequently, there will be large, short wavelength inhomogeneities in the moduli fields before thermalization, and the impact of this on their dynamics has not yet been properly explored. 

More generally, these preliminary results show that within  a complete theory of particle physics  (in this case SUSY), understanding the origin of the present-day dark matter abundance can constrain the expansion history of the universe during the primordial dark age, and lead to more precise predictions for the primordial power spectrum. 

{\it Note Added:}  While this paper was in its final stages we received a draft from the authors of \cite{Fan:2013faa}. 
In their paper they perform a comprehensive study of the nonthermal wino, performing a careful analysis which takes into account astrophysical uncertainties associated with indirect detection and additional data from HESS \cite{Abramowski:2011hc}.
In some instances they are able to arrive at more stringent constraints on the wino self-annihilation cross-section.  This should lead to an improvement in the theoretical priors used for our analysis here and so stronger constraints on inflationary model building.  

\section*{Acknowledgments}
We would like to thank Dan Hooper, Jiji Fan, Doddy Marsh, and Matt Reece for useful discussions and correspondence.
We would also like to thank Dragan Huterer and Stefano Profumo for comments on an earlier draft of this paper.
The work of OO and SW is supported in part by NASA Astrophysics Theory Grant NNH12ZDA001N, DOE grant number DE-FG02-85ER40237,
and by the National Science Foundation under Grant No. NSF PHY11-25915.
RG was partially supported by the NSF EAPSI program Grant No. 1310726, and would like to thanks the University of Auckland for hospitality.  
SW would also like to thank the DAMTP, Cambridge University, the KITP -- Santa Barbara, and the Mitchell Institute for Fundamental Physics and Astronomy for hospitality.

\bibliographystyle{unsrt}

\end{document}